# Discover Your Competition in LTE: Client-Based Passive Data Rate Prediction by Machine Learning

Robert Falkenberg, Karsten Heimann and Christian Wietfeld
Communication Networks Institute
TU Dortmund University
44227 Dortmund, Germany
Email: {Robert.Falkenberg, Karsten.Heimann, Christian.Wietfeld}@tu-dortmund.de

*Abstract*—To receive the highest possible data rate or/and the most reliable connection, the User Equipment (UE) may want to choose between different networks. However, current LTE and LTE-Advanced mobile networks do not supply the UE with an explicit indicator about the currently achievable data rate. For this reason, the mobile device will only see what it obtains from the network once it actively sends data. A passive estimation in advance is therefore not doable without further effort. Although the device can identify its current radio conditions based on the received signal strength and quality, it has no information about the cell's traffic load caused by other users. To close this gap we present an Enhanced Client-based Control-Channel Analysis for Connectivity Estimation (E-C$^3$ACE), which uncovers the cell load broken down by each single user. Based on this information and in conjunction with existing indicators like Reference Signal Received Power (RSRP) and Reference Signal Received Quality (RSRQ), a neural network is trained to perform a data rate prediction for the current LTE link. Compared to an earlier work, our approach reduces the average prediction error below one third. Applied in public networks, the predicted data rate differs by less than 1.5 Mbit/s in 93% of cases.

## I. INTRODUCTION

Mobile communication networks, especially Long Term Evolution (LTE), are attractive candidates for connecting numerous Internet of Things (IoT) devices to the Internet. Due to its high reliability, high capacity, and low latency, LTE is also a worthwhile technology to interconnect smart components and machines in upcoming *Industry 4.0* facilities as well. For example, in a future logistics storage smart containers might attach to the local available public network and enable self-reporting of the entire inventory, paging for goods, and further self-organizational procedures.

However, with a growing degree of esteem and an increasing number of devices, the network may quickly reach a congested condition and impair the mentioned performance indicators. Transmissions would be prolonged and increase the power consumption of the User Equipment (UE) due to a longer active time of the radio [1]. However, having an estimate of the currently expected data rate would enable opportunistic transmissions by postponing non time-critical transfers until the network's load relaxes. If this estimation covered the data rates of multiple LTE networks from different operators, it would even allow a dynamic switch to the currently best network.

Another application for data-rate estimation is offloading of computations from mobile devices to cloud services in order to save battery lifetime. The decision, whether it requires more energy to exchhange data with the cloud or to process the data locally, implies a qualified estimate of the radio link. This arises the need for a quick responding estimation of the current connectivity without a long-term observation of the radio link. Inducing traffic would unnecessarily stress the network and waste energy of the UE. Hence, the data rate estimation requires a passive approach instead of just benchmarking the connection with an actual transmission.

As LTE does not provide reliable indicators to readily perform such a data-rate prediction, we propose in this paper a new passive data-rate prediction system for LTE networks as illustrated in Fig. 1. It incorporates machine learning to adapt to the serving evolved NodeB (eNodeB) and the local cell environment. A specialized sniffer performs a real-time control-channel analysis to recover the set of Radio Network Temporary Identifiers (RNTIs) from every active UE in the cell. With the help of this set, it provides detailed load information of the usually unknown background traffic to a data-rate predictor. The downlink control channels are transmitted by the eNodeB and supply all active UEs with Resource Block (RB) assignments for both, downlink and uplink transmissions. Therefore, hidden stations, which are attached to the same cell and are not in range of the sniffer, have no impact on the reliability of this approach since the sniffer needs only to listen to the base station.

The predictor, a neural network, is trained in an learning cycle by repeatedly activating a Device Under Test (DUT) and measuring the actually achieved data rate in the current load situation. After finishing the training, the sniffer and the predictor are activated only if the DUT demands for a data rate estimation.

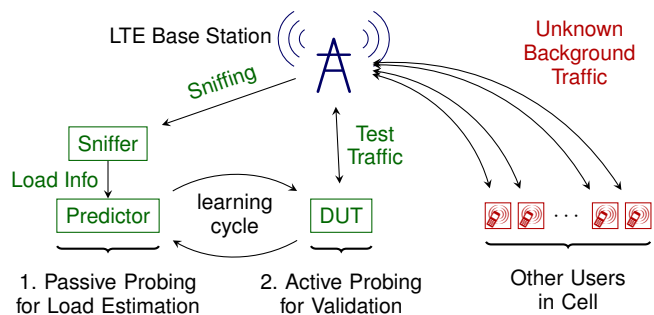

Fig. 1. Overview of the proposed prediction system in an LTE cell. Based on captured UEs and their resource demands, a predictor forecasts the data rate of the Device Under Test (DUT). Afterwards, the actual data rate of the DUT is feed back to the predictor in a learning cycle.

The remainder of this paper is structured as follows: First we evaluate different approaches for load and data rate estimation of LTE networks in Sec. II and explain the challenges of control channel analysis in Sec. III. Afterwards, in Sec. IV we capture training data in a controlled environment attached to a dedicated LTE cell. Based on this data, we train a neural network to predict the data rate of a pending transmission. Finally, in Sec. V we evaluate the performance of our predictor in the dedicated cell environment and in a public LTE network.

## II. RELATED WORK

A common approach for performance evaluation of a mobile network connection is the usage of active probing mechanisms on Commercial Off-the-Shelf (COTS) devices [2]. This requires dedicated traffic or an actual transmission to measure performance indicators like, e.g., throughput and latency. In contrast to that, passive approaches try to forecast the data rate in advance and without an active transmission based on available quality indicators in the UE like Reference Signal Received Power (RSRP) and Reference Signal Received Quality (RSRQ) [3]. However, this estimation is only reliable at low cell loads, where the impact of resource sharing among other users can be neglected. Otherwise, the provided data rate estimation can only serve as an expected upper bound for a pending transmission.

Providing more detailed information from lower protocol layers is addressed by *MobileInsight* [4]. The authors present a user space service which extracts additional data from the LTE chipset and performs, e.g., a detailed traceback of internal operating-state transitions. While this approach enables an analysis of the network behavior from the sight of a single device, it still cannot retrieve information about other users and the current cell load from the lowest protocol layers.

In contrast to COTS devices, Software-Defined Radios (SDRs) like *OpenAirInterface* [5], *srs-LTE* [6] and *openLTE* [7] provide the deepest insight into the protocol stack and allow a quick and low-cost development of new features, which may even be included in future COTS UE. With *piStream* [8] the authors perform a rate adaption of video streams to avoid video stalling by measuring the RB utilization of the attached LTE cell.

An additional breakdown to single cell users is performed with *LTEye* [9]. The authors present a non-commercial offline analysis tool for, e.g., network operators to discover inefficient radio spectrum utilization. It extracts the Physical Downlink Control Channel (PDCCH), which contains all user-specific downlink and uplink RB allocations and therefore uncovers the entire resource utilization of the regarded cell. Usually, only the receiver can verify the integrity of this data, because the attached checksum of each data structure is scrambled (via bitwise XOR) with the individual RNTI of the addressed device. *LTEye* exploits this checksum in an opposite way: Assuming no error, it blind-decodes each Downlink Control Information (DCI) candidate, computes the checksum of the structure, and recovers the unknown RNTI from the computed and the attached value. Afterwards it checks the integrity by re-encoding the DCI and comparing the resulting bits with the initially received bits before decoding. Though, as shown in [8] and [10], at non-ideal radio conditions this approach either withdraws a too high fraction of valid data or leads to numerous false-positive DCI when accepting more bit errors in the integrity check.

To overcome this insufficiency, the authors of [11] present *OWL*, a real-time capable online approach which keeps track of occurring random access procedures. These are performed when a UE wakes up from idle mode for a pending transmission, if it enters the cell after a handover, or as a response to paging. The exchanged data-structures contain the initial RNTI assignments which allow maintaining a list of active RNTIs together with a reliable Cyclic Redundancy Check (CRC) validation of DCI when decoding the PDCCH. As an assigned RNTI is released when a UE falls back into idle mode, the list will gradually reach completeness, hence being very reliable for long-term or continuous observations. However, during initialization and therefore for short-term analysis it falls back to the aforementioned re-encoding approach with its limitations at non-ideal radio conditions.

In a recent work we presented Client-based Control Channel Analysis for Connectivity Estimation ($C^3ACE$) [10], a real-time data rate estimation based on DCI analysis and the gained number of currently active users. These also are detected by blind-decoding the entire PDCCH and recovering the users' RNTIs from the checksum. In contrast to *LTEye* we keep a histogram of recent appearing RNTIs and accept only those DCI whose corresponding RNTI exceeds a small threshold in the histogram. This brings back the validation functionality of the checksum. With the risk of missing a small number of DCI from low-active users, this approach does neither rely on perfect radio conditions, nor on catching the random access procedures.

## III. CONTROL CHANNEL ANALYSIS

In an LTE cell the eNodeB, as a central authority, shares the available RBs among the active users according to its internal scheduling strategy. It transmits these resource assignments, also designated as DCI, to the UE in the PDCCH at the beginning of every LTE subframe (once per millisecond). Although this channel is not encrypted and carries all allocations for the current subframe, a normal UE is capable of decoding only its own DCI since this procedure requires the knowledge of the individually assigned RNTI. Therefore, todays LTE devices are capable to decode only their own allocations and those, which contain resources for broadcast information (e.g., system information blocks). The remaining DCI are usually ignored by the UE, since they have no relevance for a regular operation. Yet, this information can be retrieved by analyzing the control channels with the help of software defined radios, as we will present in this section. The chosen approach is intentionally kept as simple and as low in complexity as possible to allow an integration in future LTE devices.

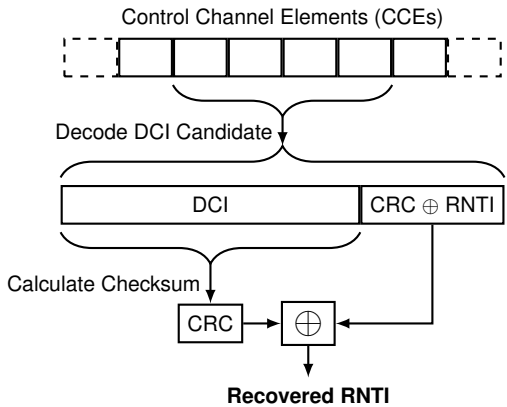

Fig. 2. Overview of PDCCH blind-decoding procedure and RNTI recovering.

### A. Structure of the Channel

The PDCCH is transmitted by the eNodeB at the start of every LTE subframe. It contains DCI for the current subframe $n$ in downlink direction and resource allocations for subframe $n+4$ in uplink direction. The delay of 4 subframes allows the UE to encode and prepare an appropriate amount of payload according to the provided resources, whereas in downlink direction, the UE can immediately decode the particular RBs from the current subframe.

From a logical point of view, the PDCCH is divided in a set of equally sized Control Channel Elements (CCEs), as shown in Fig. 2. These CCEs carry the encoded DCI structures. According to the selected *aggregation level L*, a single DCI structure may occupy 1, 2, 4, or 8 consecutive CCEs to increase its robustness against distortions. However, if a UE detects signal power on several consecutive CCEs it cannot detect the number of contained DCI structures. Instead it blind decodes those CCEs at different aggregation levels and validates their integrity by calculating and comparing the CRC checksum. Since the appended checksum is scrambled with the RNTI of the addressed UE, the UE accepts only those results, which contain its assigned RNTI as checksum.

In addition, the LTE standard defines multiple data-structure formats for DCI [12]. Depending on the direction (uplink or downlink) and the transmission mode (e.g., SISO or MIMO transmission) the DCI data structures differ either in size or by a specific bit flag setting. The rate-adaption algorithm of the data encoder fits these structs into the particular number of $L$ CCEs. Similarly to the aggregation level, a UE cannot determine the data format of an encoded DCI without decoding it. However, since decoding requires knowledge about the length of the actually contained information, a UE just blind decodes each candidate for the particular lengths of the possible formats and validates the results by their checksum. Whereas a UE can limit the number of relevant formats based on its configured transmission mode, a sniffer is obliged to incorporate all possibilities, since a eNodeB may simultaneously use multiple different DCI formats conforming to the diverse capabilities of the attached UEs.

### B. Implementation Platform

In order to build our sniffer, we modified a software defined radio implementation of an LTE UE. It is based on OPENAIR-INTERFACE's (OAI) [13] application LTE-SOFTMODEM and uses an Ettus Research USRP B210 radio front end. The sniffer hardware is shown on the right side of Fig. 4. We modified the functionality of the UE in that way, that it first synchronizes on an LTE cell at a given frequency but performs no attach request to the network. Instead, it continuously decodes the PDCCH and searches for valid DCI in every subframe. It supports DCI formats 0, 1, 1A, 1C, 2, and 2A, which covers transmission modes 2, 3, and 4 [12]. The recovered resource utilization is logged to a file or printed on the screen in real time.

### C. DCI validation

Since a UE or a sniffer cannot determine the format and aggregation level of DCI structure from the encoded raw signal (cf. Sec. III-A), the sniffer is obliged to perform a blind decoding of the PDCCH for all possible combinations and validate the decoded output afterwards. Whereas a UE performs this validation by matching the candidate's checksum to its assigned RNTI, a sniffer initially has no knowledge about the current set of valid RNTIs in the cell. Therefore, the sniffer must first reconstruct this RNTI set from the blind decoded DCI by recovering the RNTIs from the checksums. As most of the blind decoded DCI candidates contain false information, due to a format or aggregation level mismatch, we performed a two-step filtering of the DCI candidates:

First, the eNodeB may place DCI for a particular RNTI only in a subset of the available CCEs [12]. This subset is derived from the RNTI and the current subframe index $0\ldots9$ and reduces the blind-decoding complexity of a regular UE by factor 5 in 10 MHz cells and by factor 10 in 20 MHz cells. Consequently, if the sniffer resolves a DCI to an RNTI, which is not allowed on this position, this DCI will be dropped without any further processing.

The second step counts the recurrence frequency of every remaining RNTI (16 bit integer each) in a histogram, regardless of their validity as shown in Fig. 3. If the frequency of an RNTI exceeds a predefined threshold, it is assumed as true and the corresponding DCI passes the validity check. As we already showed in [10], a histogram with a time window of 500 ms and a threshold value of 8 leads to a probability of $5.6 \cdot 10^{-7}$ per subframe for classifying a false-positive RNTI. Due to the fact that each UE, together with its RNTI, is configured to a fixed transmission mode, each RNTI is bound to a particular DCI format in the downlink. Therefore, we implemented a dedicated history for each DCI format, which prevents a *crosstalking* of occurrences between different formats.

### IV. ACQUISITION OF TRAINING DATA

This section describes our measurement setup and the process of data capturing. Afterwards, we describe the pre-processing step for feature generation and the application of machine learning to train a predictor on the captured data.

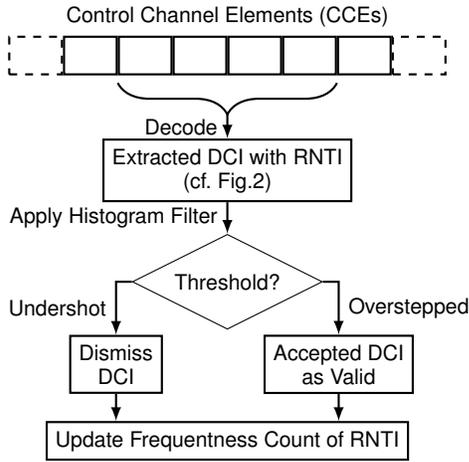

Fig. 3. Operating principle of the histogram filter.

TABLE I
PROPERTIES OF THE MEASUREMENT SETUP

| LTE Cell Properties | |
|---|---|
| Frequency Band | 2.6 GHz (Band 7) |
| EIRP | 631 mW (28 dBm) |
| Bandwidth | 10 MHz (50 RBs) |
| Duplex Mode | Frequency Division Duplex |
| **Device Under Test Properties** | |
| RSRP (*Near*) | −80 dBm |
| RSRP (*Distributed*) | −93 dBm |
| RSRP (*Far*) | −101 dBm |

### A. Environment Setup

For our measurements, we used a dedicated LTE network with a configuration as listed in the upper part of Tab. I. During our experiments, no other interfering networks were in range to disturb the signal. As shown in Fig. 4, the base station was placed in our office environment, as well as several Smart Traffic Generators (STGs) [14], the DUT, and the sniffer. The sniffer is a software defined radio, which runs the modified software of OPENAIRINTERFACE's UE on a compact general-purpose computer and writes all captured control-channel data into a log file. It is shown on the right side of the figure. In contrast, the STGs and the DUT are small embedded platforms equipped with an LTE modem, which can be configured to induct versatile traffic patterns into the network.

The STGs and the DUT were placed in three different arrangements (*distributed, near, far*) as shown in Fig. 4. This diversity shall prevent the machine learning algorithm from adapting too close to a specific setup, which would impair the general applicability of the predictor. While the *distributed* scenario represents a typical spreading of UEs in a mobile network, the predictor should also be prepared for situations, where the UEs are mostly placed *near* or *far* away from the base station. This represents large rural cells and small urban cells, respectively. Furthermore, each of these three screnarios includes measurements, where the DUT was placed at *near*, *far*, and *intermediate* distance to the eNodeB.

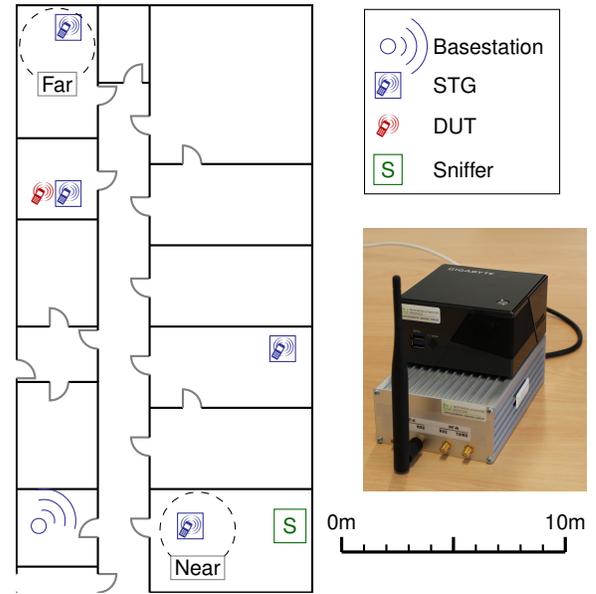

Fig. 4. Photography of the sniffer and overview of the measurement setup for training in the *distributed* scenario. The other scenarios, *near* and *far*, cumulate all STGs inside the marked areas near and far, respectively. In addition, each scenario includes measurements where the DUT was also placed at the *near* and *far* position.

For each arrangement we used three different types of background traffic by the STGs and performed measurements for both, downlink and uplink direction:

- High load: continuous best effort Transmission Control Protocol (TCP) transmission to stress the network.
- Low load: continuous User Datagram Protocol (UDP) transmission with limited data rate to 64 kbit/s in downlink and 16 kbit/s in uplink direction.
- Mixed load: each STG performs every 30 s a UDP transmission for randomly 5 s, 10 s, or 25 s. Each transmission is performed at a randomly chosen data rate out of 0.064 Mbit/s, 0.5 Mbit/s, 1 Mbit/s, 3 Mbit/s, 5 Mbit/s, and 10 Mbit/s for downlink and 0.016 Mbit/s, 0.5 Mbit/s, 1 Mbit/s, 2 Mbit/s, 4 Mbit/s, and 10 Mbit/s for uplink.

In addition, the STGs randomly detached and attached to the eNodeB, which simulates the fluctuating number of active UEs in an LTE cell.

### B. Capturing, Preprocessing and Feature Generation

As explained in the previous subsection, numerous scenarios were covered by the environment setup to collect versatile training data for the machine learning algorithm. Each scenario includes approximately 160 transmissions of the DUT, consisting of rotary downloads and uploads of an 1 MB file over the LTE link. For every transmission the DUT writes the achieved average data rate $r$, the initial RSRP and RSRQ values, and a timestamp into a log file. Each transmission is followed by a resting period of 10 s.

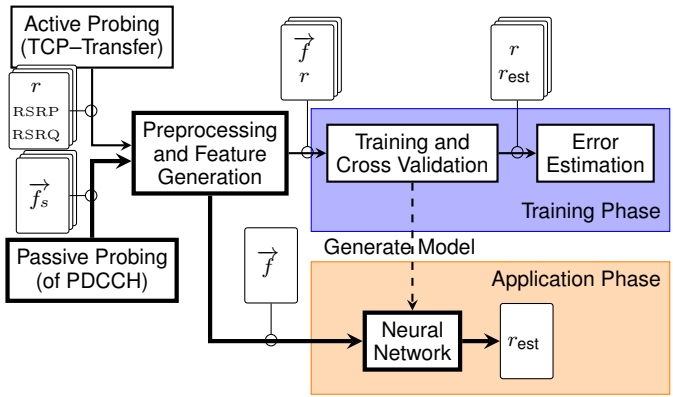

Fig. 5. Overview of the processing pipeline and the training process of the data-rate predictor. After training only the bold components are required for the data-rate prediction.

At the same time, the sniffer continuously logs and time-stamps the following values from the control channel analysis:
- RNTI: identifying a user
- $N_{\text{RBs}}$: number of allocated RBs
- Modulation and Coding Scheme (MCS): representing the coding rate
- Transport Block Size (TBS): amount of data
- Direction (downlink or uplink)
- $N_{\text{UE}}$: number of currently active RNTIs from histogram

The values have a granularity of one LTE subframe, which corresponds to 1 ms. We combine them to vector $\vec{f_s}$ of input features from the sniffer.

Based on the synchronized timestamps data from both log files is being merged and preprocessed to generate labeled feature vectors for the machine learning algorithm (cf. Fig. 5). For every transmission of the DUT the preprocessing step calculates average values and standard deviation of $N_{\text{RBs}}$, $N_{\text{UE}}$, TBS, and MCS over a period of 1 s before the actual DUT transmission. These values, together with RSRP and RSRQ from the DUT, form a single feature vector $\vec{f}$ for the machine learning algorithm. Each vector is labeled by the achieved average data rate $r$ during the particular transmission.

### C. Machine Learning Approach

Based on the captured data, we trained an artificial neural network to perform the data rate prediction by using the machine learning software *RapidMiner* [15]. We setup *RapidMiner* to train a neural network and performed a cross validation on the given data set, as shown in Fig. 5. While we also tried different configurations, a neural network with two hidden layers of 10 and 5 neurons, gave us the best results. Other parameters, such as learning rate and momentum, were optimized by an evolutionary algorithm, which is also provided by the software. A cross validation, in contrast to a split validation, returns an estimate $r_{\text{est}}$ for every input vector and hence increases the number of ratable data.

After completing the training, the generated model can be exported and used in the application phase to perform the data rate prediction. In this phase the complexity of the predictor narrows to the bold components in Fig. 5. Only if the predictor is moved to another cell, the learning cycle should be repeated (cf. Fig. 1), since operators might configure the eNodeB's scheduler differently for distinct environments.

## V. EVALUATION

In order to evaluate the accuracy of the proposed data-rate predictor, we performed numerous measurements in our dedicated LTE network for three types of cell traffic, as explained in Sec. IV. In addition, we also measured the performance of the predictor in a field test attached to a public network. Each prediction consists of an estimated data rate $r_{\text{est}}$ by the predictor and the actually achieved data rate $r$ in a subsequent transmission, as shown in Fig. 5. For comparison, each sample also contains a data rate estimation based on the earlier approach $C^3$ACE [10], which is derived only from the average number of detected active UEs over 1 s and the average data rate in an empty cell. Its data rate estimate is calculated as follows:

$$r_{\text{est}}^{C^3\text{ACE}} = \frac{1}{N+1} \cdot r_0(\text{MCS}_{\text{UE}}), \quad (1)$$

where $N$ describes the detected number of active users and $r_0(\cdot)$ is the average data rate of the UE according to its current radio condition and the corresponding MCS in an empty cell. In contrast to $C^3$ACE, the proposed Enhanced Client-based Control Channel Analysis for Connectivity Estimation (E-$C^3$ACE) utilizes the full feature set described in Sec. IV-B and a neural network for its estimation.

### A. Data Rate Prediction in a Dedicated Network

In this section we present the performance evaluation after the training phase in a dedicated LTE network, as described in Sec. IV. Fig. 6 shows the predicted data rate by the proposed E-$C^3$ACE (green) in the full-load scenario, in which the active UEs continuously transmit as much data as possible. Since each prediction process is followed by a transmission of the DUT, the actually achieved data rate is plotted as black samples for a ground-truth reference. Samples from both sets lie very close to each other, which reflects a very high prediction accuracy. Although the figure carries out only a single input dimension, which is the detected number of active devices averaged over 1 s, E-$C^3$ACE still considers numerous additional features for its prediction (cf. Sec. IV-B). Yet, the one dimensional representation gives a good comparison to predictions by the earlier approach $C^3$ACE (plotted in red). While the red curve uses the average data rate in an empty cell as a reference, the blue curve marks the upper bound for the achievable $C^3$ACE predictions, by choosing the reference data rate retrospectively to minimize the overall Root Mean Square Error (RMSE). Therefore, this shows the theoretical limit of the $C^3$ACE approach.

Generally, $C^3$ACE and its upper bound tend to underestimate the data rate, thus performing to pessimistic estimations. In contrast, the proposed E-$C^3$ACE estimates has no general tendency to under- or overestimate the data rate. Due to machine learning it adapts to the eNodeB's scheduler, which might slightly reduce the robustness of modulation and coding at higher loads to countersteer congestion. The overall RMSE of 356 kbit/s is much lower than 1717 kbit/s and 1271 kbit/s, which are made by C³ACE and its upper bound, respectively. In comparison to C³ACE, E-C³ACE reduces the prediction error to 28 %. The improvement of E-C³ACE becomes even clearer when comparing the error's Empirical Cumulative Distribution Function (ECDF) of both approaches, (cf. Fig. 7). In 99 % of cases the data-rate estimate error by E-C³ACE is smaller than 1.5 Mbit/s, while even the upper bound of C³ACE undershoots this mark only in 85 % of cases. Actual C³ACE predictions achieve this accuracy only in 70 %.

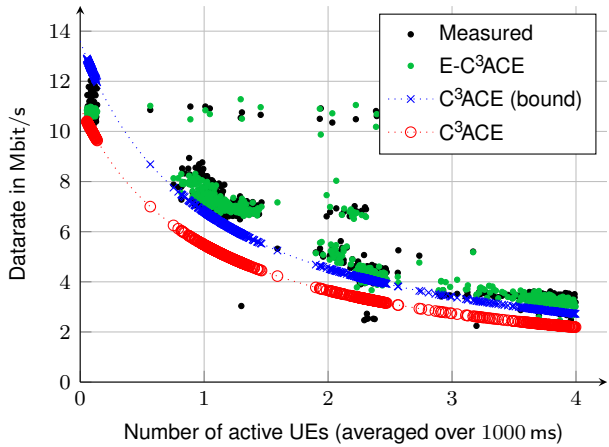

Fig. 6. Predicted data rate by C³ACE and the proposed E-C³ACE in relation to the number of active UEs in a dedicated LTE cell at high load. While C³ACE mostly underestimates the data rate, E-C³ACE hits the actually measured rate more precisely.

TABLE II
PREDICTION ERROR (RMSE) IN DIFFERENT LOAD SCENARIOS

| Scenario | E-C³ACE | C³ACE (bound) | C³ACE |
|---|---|---|---|
| High Load | 356 kbit/s | 1271 kbit/s | 1717 kbit/s |
| Mixed Load | 1078 kbit/s | 6546 kbit/s | 8881 kbit/s |
| Low Load | 1534 kbit/s | 2894 kbit/s | 3179 kbit/s |

The results of the remaining scenarios are listed in Tab. II. In case of mixed cell load, the RMSE of E-C³ACE triples as a consequence of higher dynamics in the network. For example, the DUT may achieve a higher data rate than expected if another device terminates its activity during the DUT's transmission. At low cell loads, on the other hand, the DUT mostly achieves its maximum data rate. However, the TCP congestion control mechanism greatly decelerates the transmission, if the data rate momentary falls due to short buffered bursts by other participants. Therefore, the largest RMSE appears in these scenarios as a consequence of highest data rates and largest data rate drops.

### B. Data Rate Prediction in a Public Network

In order to evaluate the performance of our approach in an active public network, we performed field measurements in a rural environment attached to a 800 MHz cell of *Deutsche*

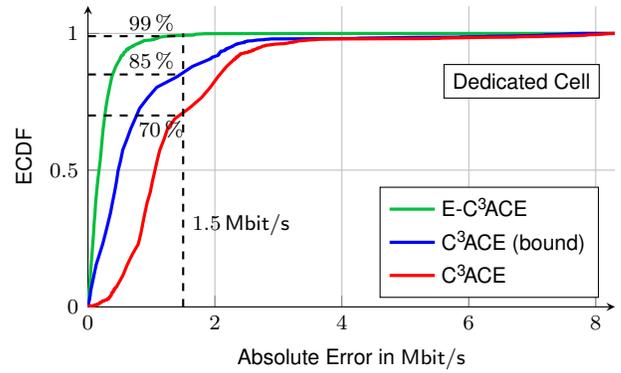

Fig. 7. ECDF of the absolute prediction error in the high load scenario. Compared to C³ACE the proposed E-C³ACE has a significantly higher accuracy with an error $< 1.5\,\mathrm{Mbit/s}$ in 99 % of cases.

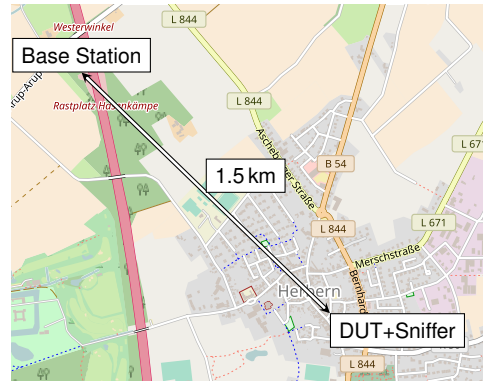

Fig. 8. Overview of the data rate prediction setup in a public LTE network. Sniffer and the DUT were placed at a distance of 1.5 km to the next LTE base station. (Map: ©OpenStreetMap contributors, CC BY-SA)

*Telekom AG* in Germany. The eNodeB is located on a parking space at the A1 motor highway next to the small village Ascheberg-Herbern. DUT and the sniffer were placed indoor at a non-line-of-sight position with a distance of 1.5 km to the eNodeB. Our measurements cover a time interval of 27 hours with one data sample every 46.44 seconds on average.

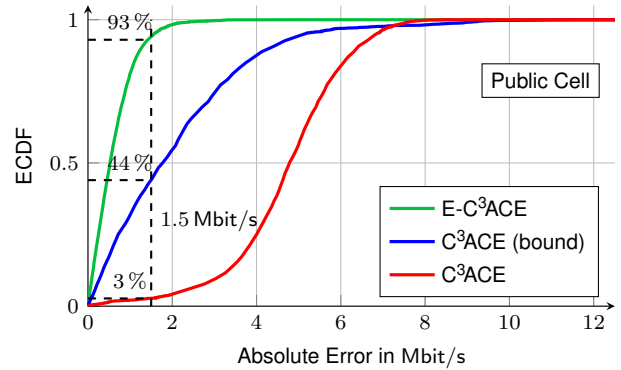

Fig. 9. ECDF of the absolute prediction error in a public network. E-C³ACE still has a significantly higher accuracy than C³ACE and an error $< 1.5\,\mathrm{Mbit/s}$ in 93 % of cases.

Fig. 9 shows the performance of the proposed predictor in the field test. It still achieves in 93 % of cases an error below 1.5 Mbit/s. The overall RMSE doubles to 0.833 Mbit/s, if comparing to the 0.411 Mbit/s from laboratory predictions. Comparing the performance to the earlier C³ACE approach shows great benefits of incorporating multiple indicators for the data rate estimation. Here, C³ACE achieves only in 41 % an error below 1.5 Mbit/s using a-posteriori knowledge.

## VI. CONCLUSION

In this paper we proposed a passive approach for forecasting the data rate in User Equipment (UE), which is attached to an LTE mobile network. Although LTE does not readily provide this information to its devices, a prior knowledge of the expected data rate would for example help mobile devices avoiding congestions by delaying transmissions to uncongested moments. This could save battery lifetime since the transmissions would terminate faster.

To close this gap, we perform a client-based control channel analysis and uncover the eNodeB's entire resource distribution among all active participants. We showed in an earlier work (C³ACE) [10] that a coarse grained data rate estimation is already possible by dividing the maximum achievable data rate of a UE by the number of detected active cell users. This work extends that approach (E-C³ACE) by incorporating machine learning to train an artificial neural network and perform more accurate predictions based on numerous additional input parameters. For this purpose, we first extended the control channel sniffer to provide information such as number of allocated RBs per users and the particularly assigned MCS. In addition, we also capture readily available quality indicators at the Device Under Test (DUT), such as $RSRP$ and $RSRQ$, to consider the current radio link quality and derive the maximum achievable data rate of the device. Secondly, we performed excessive measurement campaigns in a dedicated LTE test network to capture the necessary training data for the neural network. We used different traffic patterns, varied the number of traffic generators, and placed the DUT at different locations to cover a broad range of possible scenarios. The training process incorporates a cross validation and therefore utilizes the entire data set for generating the model instead of holding back a subset for a split validation.

The evaluation of the proposed E-C³ACE shows at high cell loads an increase of accuracy by more than factor 3 compared to the earlier approach C³ACE. While C³ACE predictions differ from truth by less than 1.5 Mbit/s only in 70 % of cases, the proposed E-C³ACE undershoots this error mark in 99 %. In a second campaign we challenged our approach in a public mobile network. Again, we achieved a considerably higher prediction accuracy compared to C³ACE. The proposed E-C³ACE induced a prediction error of less than 1.5 Mbit/s in still 93 % of cases.

In future works we will apply our approach in even more challenging environments, such as train stations. Furthermore, we will train the neural network across multiple base stations and environments in order to generate a more generic predictor which is not bound to a single cell. This will untighten requirements for a learning cycle but requires a significantly increased number of collected training data to incorporate the environment.

## ACKNOWLEDGMENT

Part of the work on this paper has been supported by Deutsche Forschungsgemeinschaft (DFG) within the Collaborative Research Center SFB 876 "Providing Information by Resource-Constrained Analysis", project A4.